# A General Approach to High Efficiency Perovskite Solar Cells by Any Antisolvent

*Alexander D. Taylor,[1,2,3] Qing Sun,[1] Katelyn P. Goetz,[1,2,3] Qingzhi An,[1,2,3] Tim Schramm,[2] Yvonne Hofstetter,[1,2,3] Maximillian Litterst,[1] Fabian Paulus,[1,3] and Yana Vaynzof[1,2,3]\**

[1] Kirchhoff Institute for Physics and Centre for Advanced Materials, Ruprecht-Karls-Universität Heidelberg, Im Neuenheimer Feld 227, 69120 Heidelberg, Germany

[2] Integrated Centre for Applied Physics and Photonic Materials, Technical University of Dresden, Nöthnitzer Str. 61, 01187 Dresden, Germany

[3] Center for Advancing Electronics Dresden (cfaed), Helmholtzstraße 18, 01089 Dresden, Germany

**Corresponding Author**

*Yana Vaynzof, e-mail: yana.vaynzof@tu-dresden.de

**Abstract**

Deposition of perovskite thin films by antisolvent engineering is one of the most common methods employed in perovskite photovoltaics research. Herein, we report on a general method that allows the fabrication of highly efficient perovskite solar cells by any antisolvent via the manipulation of the antisolvent application rate. Through a detailed structural, compositional and microstructural characterization of perovskite layers fabricated by 14 different antisolvents, we identify two key factors that influence the quality of the perovskite active layer: the solubility of the organic precursors in the antisolvent and its miscibility with the host solvent(s) of the perovskite precursor solution. Depending on these two factors, each antisolvent can be utilized to produce high performance devices reaching power conversion efficiencies (PCEs) that exceed 21%. Moreover, we demonstrate that by employing the optimal antisolvent application procedure, highly efficient solar cells can be fabricated from a broad range of precursor stoichiometries, with either a significant excess or deficiency of organic iodides.

**Introduction**

Perovskites display a number of properties that directly translate to high performance in photovoltaic devices, such as low exciton binding energies,[1] long charge-carrier diffusion lengths,[2] and high absorption coefficients.[3] Such exceptional electronic behavior is tantalizing, and made more so by the low cost of their film fabrication[4]. Because they are made from earth-abundant materials and can be processed by low-temperature solution methods, perovskites have the potential to expand PV use by dramatically lowering the device payback time.[5,6] Researchers have repeatedly proven how to effectively combine these factors by simple spin-coating techniques, reporting power conversion efficiencies (PCEs) in excess of 20% and a current record of 25.2%.[7]

Among the various methods to deposit perovskite layers, such as spin coating, inkjet printing[8,9], thermal evaporation[10–12], and many others, the most popular and commonly used is the so-called 'solvent engineering' method.[13] Here the spin coating of the perovskite precursor solution employs an antisolvent treatment to facilitate the removal of the host solvent(s) and initiate crystallization of the perovskite film. Several studies have lent insight into the optimal application of this step; for example, Wang *et al.* observed that it must be timed to take place within the so-called sol-gel phase of crystallization.[14] Other parameters of antisolvent application have also been investigated, including the antisolvent volume, spinning parameters, and atmosphere;[15–17] however a clear understanding of these variables is yet to emerge. Even the choice of antisolvent is not straightforward.[18] Success has been demonstrated by researchers using antisolvents with no apparent commonality in physio-chemical properties; for example, both highly polar (such as ethyl acetate and isopropyl alcohol)[19] and nonpolar solvents (toluene)[13] have been used to form high-performance devices. For boiling point it is likewise; solvents with both extremely low (diethyl ether)[20] as well as high (chlorobenzene)[21] boiling points are commonly reported to yield high PCE devices. On the other hand, many solvents have been shown not to work well as antisolvents, leading to poor perovskite film formation and low device efficiencies.[15,22]

In this work, we explore 14 antisolvents and demonstrate a general approach to achieving high-performance triple-cation $Cs_{0.05}(MA_{0.17}FA_{0.83})_{0.95}Pb(I_{0.9}Br_{0.1})_3$ solar cells from any antisolvent. We show that by changing the duration of the antisolvent application, the device performance for certain antisolvents can be increased from non-functional to over 20% PCE. We show that antisolvents generally fall into three categories: those that favor short application times, those that are largely unaffected, and those that perform best with longer application times. By performing detailed morphological, compositional and microstructural characterization of the perovskite layers, we identify the effects of the different classes of antisolvents on the perovskite film formation. We find that the solvent categorization is related to two fundamental properties: the degree of solubility of the organic iodides in the antisolvent and the miscibility of the antisolvent with the perovskite precursor host solvent(s). Depending on these two factors, tuning the application time results in efficient photovoltaic devices from any antisolvent. Finally, we also demonstrate that by using the optimal application time, it is possible to significantly expand the range of stoichiometries that lead to high device performance, thus eliminating the need for the commonly used $PbI_2$ excess in the perovskite composition. Our results represent a tremendous step toward a fundamental understanding of the role of antisolvents in perovskite film formation and demonstrate a general approach for efficient perovskite solar cells from any antisolvent.

**Results**

**Adjusting the Duration of Antisolvent Application**

The process of fabricating perovskite films by solvent engineering is schematically shown in Figure 1a. In short, a concentrated perovskite precursor ink is deposited via spin coating, followed by the application of an antisolvent at a fixed time before the end of the spin-coating procedure. The application of the antisolvent is not instantaneous and its duration ($\Delta t$) has yet to be considered as an important factor for the perovskite film formation. To study this, a simple method to adjust the duration of antisolvent application is to employ two sizes of micropipettes, which dispense the same volume of solvent (200 µl) over different lengths of time. As is shown in Figure 1b, the 1000 µL pipette has a significantly wider tip radius than the 250 µL, meaning the height of the solvent (for a given volume) in the tip is lower. Therefore, the plunger for the 1000 µL pipette must traverse a shorter distance to extrude the same volume of solvent (200 µl) as the 250 µL pipette, leading to a faster extrusion speed. To quantify these rates and the duration of antisolvent application, we filmed example antisolvent extrusions ('fast' and 'slow'), and measured the time required for each via frame counting (supplementary section S1). This analysis resulted in approximate extrusion rates of 1100 µL/s for fast ($\Delta t = 0.18$s), and 150 µL/s for slow ($\Delta t = 1.3$s).

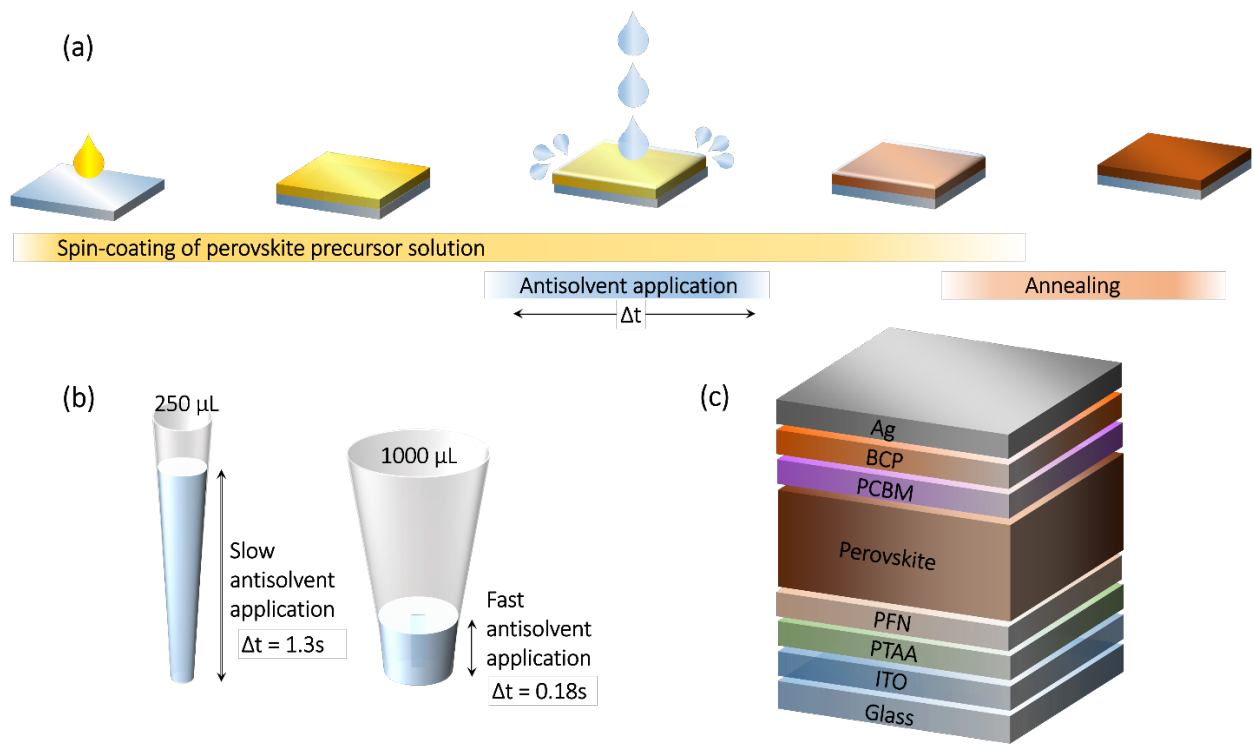

**Figure 1.** (a) Schematic depiction of the perovskite layer fabrication process. (b) Illustration of the 1000 and 250 μL pipettes that were used to adjust the duration of the antisolvent application step. For the same applied force and solvent volume (200 μl), the extrusion rate is 'fast' for the 1000 μl pipette and 'slow' for the 250 μl pipette. (c) Schematic structure of the photovoltaic devices fabricated in this work.

**The Impact on Solar Cell Performance**

To investigate the effect of adjusting the duration of antisolvent application step, we fabricated nearly 800 "triple cation" $Cs_{0.05}(MA_{0.17}FA_{0.83})_{0.95}Pb(I_{0.9}Br_{0.1})_3$ perovskite PV devices across 14 different antisolvents (see supplementary note S2, Table S1 for antisolvent properties, and Figure 1c for full device stack used), and compared their photovoltaic performance when using a fast or slow antisolvent application (called 'fast' and 'slow' devices for simplicity). The chemical structures of the antisolvents are depicted in Figure 2a, with the solvents categorized according to their photovoltaic performance. A summary of the resulting open-circuit voltage ($V_{OC}$), short-circuit current ($J_{SC}$), fill factor (FF), and PCE is shown in Figure 2b. Broadly speaking, almost every antisolvent yields devices with PCEs approaching or exceeding 20 %, with the best performers having a consistent $V_{OC}$ of ~1.1 V, a FF between 75-83 %, and a $J_{SC}$ of 22-23 mA/cm$^2$. When considering fast versus slow antisolvent application, differences between the solvents are immediately apparent. The antisolvents can be placed into three categories, types I-III, based on these differences. Type I antisolvents, consisting of the alcohol series ethyl, isopropyl, and butyl alcohols, result in better devices when the antisolvent is applied quickly. As shown in Figure 2b, fast application of the antisolvent leads to an equally high $V_{OC}$ and similar values for the $J_{SC}$, FF, and PCE, while using a slow application negatively affects the performance, from a small, but noticeable, difference for butanol to near non-functionality (<5% PCE) for ethanol. Furthermore, especially in the FF, a significantly broader distribution of values is obtained for the 'slow' devices. The type III antisolvents, on the other hand, show the exact opposite trend. Here, slow antisolvent application yields good performance with a narrow distribution, while the fast devices perform between slightly and significantly worse. Mesitylene has the most extreme difference: while slow application yields competitive performances, fast extrusion does not result in a single functional device. Concurrently, the 'fast' devices of the other antisolvents in this category contain a higher proportion of short-circuited devices. In the final category, the performances of type II antisolvents are largely unaffected by the duration of antisolvent application. Notably, for all the tested antisolvents, the highest performance devices are at roughly the same level, reaching average PCEs around 18% and champion pixels of over 21%. Example J-V curves for 'fast' and 'slow' pixels are shown in the supplementary information section S2, Figure S1.

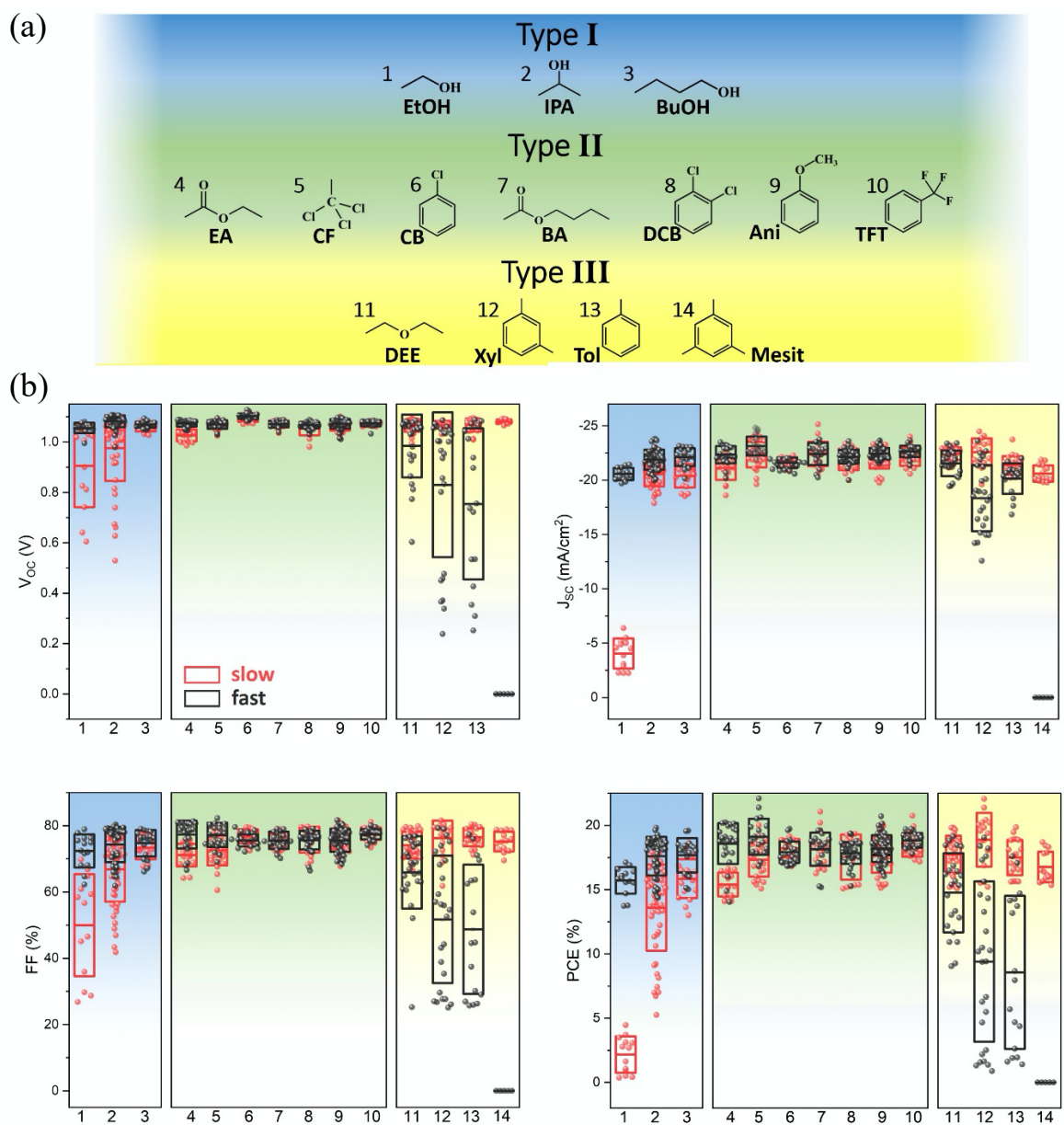

**Figure 2.** (a) The 14 antisolvents used in this experiment and the (b) photovoltaic performance of devices resulting from a 'fast' or 'slow' antisolvent application. The antisolvents are categorized as type I, II, or III according to their PV performance. The solvents are abbreviated as follows: 1: ethanol (EtOH), 2: isopropanol (IPA), 3: butyl alcohol (BuOH), 4: ethyl acetate (EA), 5: chloroform (CF), 6: chlorobenzene (CB), 7: butyl acetate (BA), 8: 1,2-dichlorobenzene (DCB), 9: anisole (Ani), 10: trifluorotoluene (TFT), 11: diethyl ether (DEE), 12: m-xylene (Xyl), 13: toluene (Tol), 14: mesitylene (Mesit).

**Type I antisolvents**

Microstructural characterization of films fabricated from type I antisolvents reveal stark differences between slow and fast antisolvent application (Figure 3). In the case of fast antisolvent application, scanning electron microscopy (SEM) images show dense and compact perovskite films with some phase-separated lead iodide (bright features in the SEM, Figures 3-5) on the surface.[23] Cross-sectional SEM confirms that the films consist of relatively large perovskite grains, which extend throughout the entire film thickness. In contrast, films formed by slow antisolvent application exhibit a far inferior microstructure. The surfaces of these films contain significantly higher amounts of lead iodide, particularly evident in the EtOH and IPA samples, and often display pinholes or small voids. Additionally, cross-sectional imaging reveals the formation of large voids at the interface with the hole-extraction layer PTAA. The apparent film quality observed via SEM aligns well with the PV results; as an example, the large density of voids observed in the slow EtOH films leads to a very poor hole extraction, greatly limiting the $J_{SC}$ and the overall photovoltaic performance. The large distribution of the photovoltaic performance of the slow IPA films is likely to be caused by the formation of pinholes and small voids in the devices' active layers. Butanol, on the other hand, shows the fewest voids at the interface with PTAA for the slow antisolvent application, consistent with the smallest difference in PV performance.

Structural characterization by 2D X-ray diffraction (2D XRD) reveals another interesting feature of films formed by type I antisolvents. While the films exhibit the characteristic perovskite diffraction patterns with a cubic unit cell with lattice constants a=b=c=6.305 Å,[24] the distribution of intensities along the Debye rings demonstrates that Type I antisolvents result in polycrystalline perovskite films with a remarkably high degree of preferred orientation.

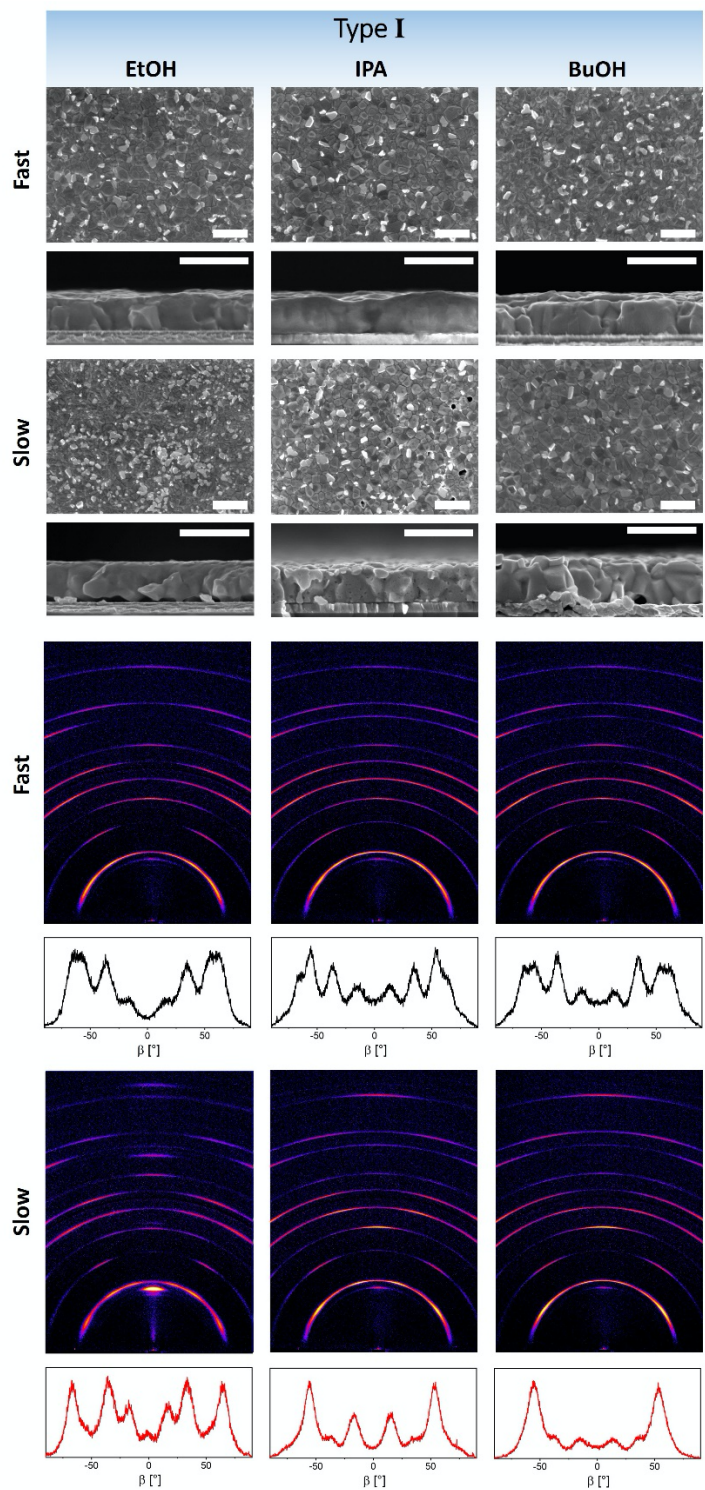

**Figure 3.** Top: surface and cross-sectional scanning electron microscopy images of perovskite films formed from type I antisolvents (EtOH, IPA and BuOH). Scale bar is 1 μm. Bottom: 2D XRD maps and corresponding ß integration of the (100) reflection to visualize changes in grain orientation.

A fast antisolvent application lowers this degree of ordering, but it is still significantly higher than that of other antisolvents and previous literature reports.[25] Additionally, the XRD measurements from all films formed by type I antisolvents show a significant contribution of lead iodide ($2\theta = 12.6°$), with EtOH displaying the largest signal, then IPA, and BuOH the smallest. Furthermore, the $PbI_2$ signal is amplified in the case of slow antisolvent application versus fast. This is in agreement with the observations by SEM and X-ray photoemission spectroscopy (supplementary note S3, Figure S2).

**Type II antisolvents**

From the seven type II antisolvents, Figure 4 shows the structural and microstructural characterization of representative three antisolvents: CF, CB and TFT, with the other four shown in the SI (supplementary note S4, Figure S3). SEM imaging reveals that all type II antisolvents result in the formation of high quality, uniform, and pinhole-free films, independent of the duration of antisolvent application. The films show a far smaller lead iodide content than type II, evident both in the surface SEM images and the 2D XRD maps. XPS measurements confirm the very similar surface composition of all films formed by type II antisolvents.

Similar to the type I antisolvents, the films exhibit perovskite diffraction patterns with the same cubic unit cell parameters, but unlike type I, no significant preferential orientation is observed for films formed with type II antisolvents, in particular those formed via a fast application. The minor differences in the degree of orientation between the antisolvent application times do not seem to correspond with differences in the photovoltaic performance of the devices.

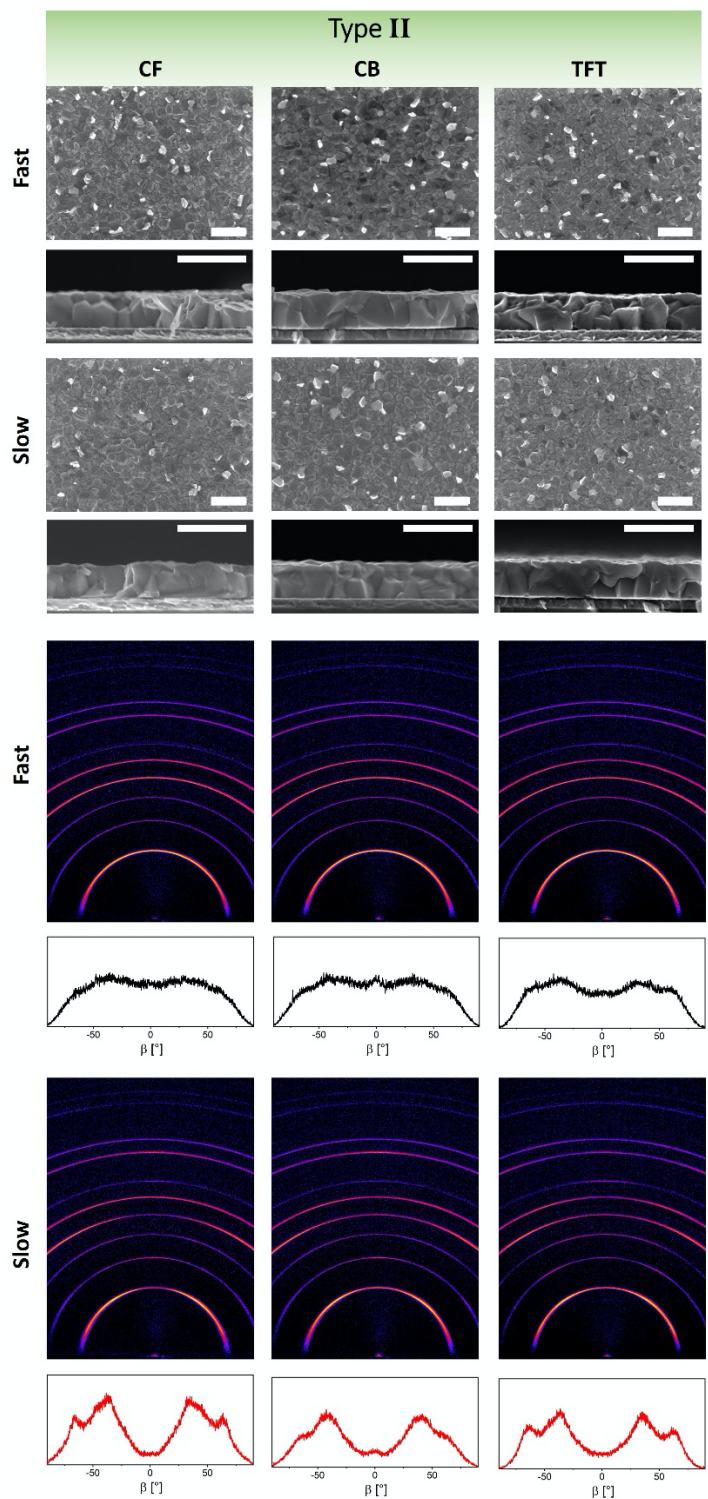

**Figure 4.** Top: surface and cross-sectional scanning electron microscopy images of perovskite films formed from selected type II antisolvents (CF, CB and TFT). Scale bar is 1 µm. Bottom: 2D

XRD maps and corresponding ß integration of the (100) reflection to visualize changes in grain orientation.

**Type III antisolvents**

The differences between films formed by slow and fast application of type III antisolvents are easily visible by eye (supplementary note S5, Figure S4). Films deposited via a fast antisolvent application result in only a portion of the film – corresponding to the area where the antisolvent is dispensed – appearing dark and shiny, apparently indicating the formation of the desired perovskite phase. Other areas, however, take on a hazy appearance, or do not change to the black perovskite phase at all. This is made apparent by optical transmission microscopy: here, the sample shows a clear boundary between the area where the black perovskite phase forms and the area where it remains unconverted (supplementary note S5, Figure S5). Interestingly, SEM imaging reveals that this central perovskite region appears similar to the other high-performing films (Figure 5 and supplementary note S6, Figure S6), apart from toluene which exhibits a significant number of pinholes. Nevertheless, 2D XRD confirms that the central region is a polycrystalline perovskite with no significant preferential orientation. It is likely that this partial conversion is a result of the central spot, where the antisolvent is dispensed, being exposed to the antisolvent for a longer amount of time than the surrounding areas, thus nearing the conditions of the slow application. The varying degrees of overlap of this central perovskite area with the device's electrodes lead to a large distribution of photovoltaic performance with significant losses due to shunts via the unconverted areas.

In contrast, slow application of type III antisolvents results in compact and uniform films with a smaller overall amount of residual $PbI_2$ (Figure 5). These films are very similar in microstructure and crystalline orientation to those formed by type II antisolvents, with many devices showing photovoltaic performance with efficiencies surpassing 20%. This is especially noteworthy in the case of mesitylene, where fast application of mesitylene leads to particularly poor films, resulting in no functional photovoltaic devices.

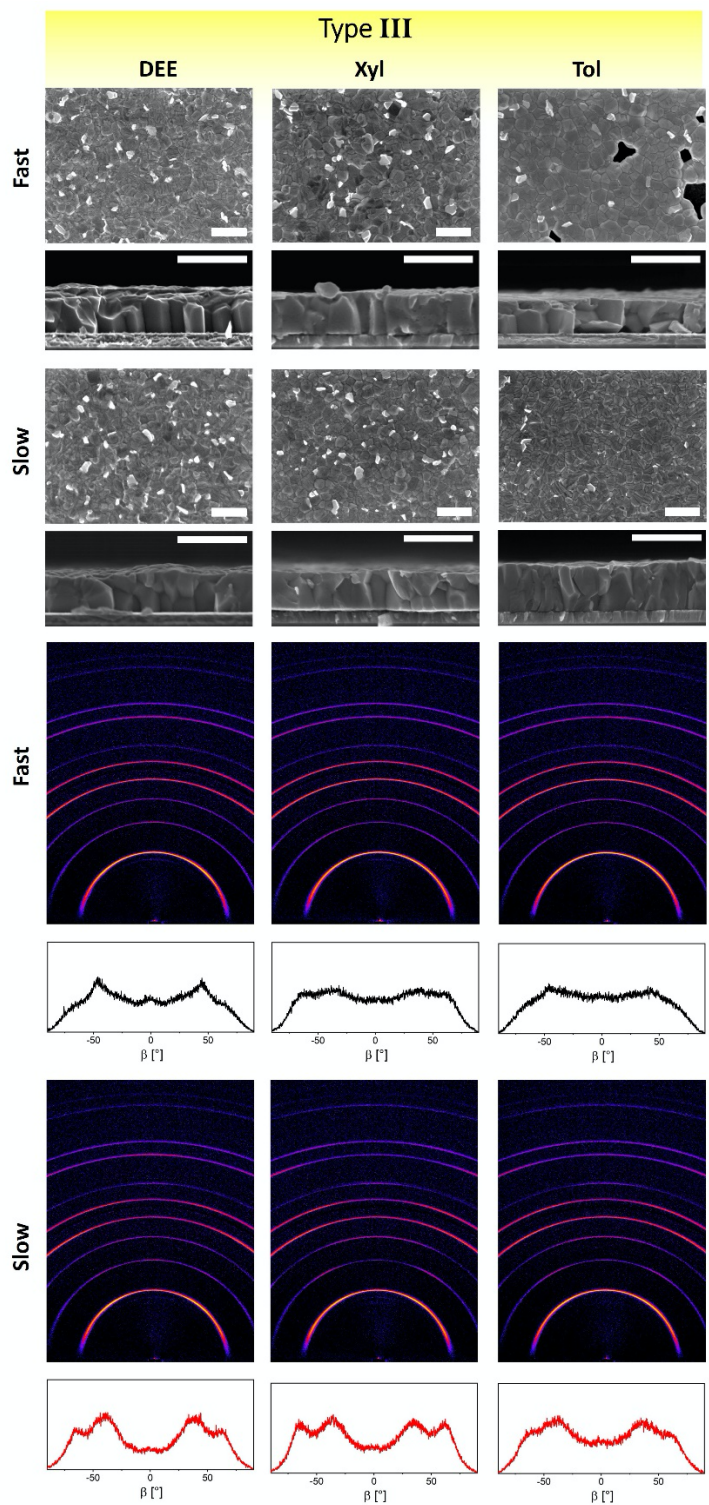

**Figure 5.** Top: surface and cross-sectional scanning electron microscopy images of perovskite films formed from selected type III antisolvents (DEE, Xyl and Tol). Scale bar is 1 μm. Bottom: 2D XRD maps and corresponding ß integration of the (100) reflection to visualize changes in grain orientation.

**Origin of Antisolvent Categorization**

The significantly damaged microstructures of slow type I and fast type III, in conjunction with altered ratios of phase-separated PbI$_2$, suggests that the relative solubility of the precursor components in the various antisolvents plays a key role in film formation, and that by changing the duration of the antisolvent application speed, one can tune this effect.[26] To explore this, we prepared 2 M methylammonium iodide (MAI) or formamidinium iodide (FAI) solutions in 4:1 dimethylformamide:dimethyl sulfoxide (DMF:DMSO), and tested the solubility/miscibility in each antisolvent. The volume ratio was chosen to be 6:1 antisolvent:host, as this is approximately the volume ratio used in the perovskite fabrication process. The results of this test for MAI are shown in Figure 6, and for FAI are shown in supplementary note S7, Figure S7a.

While the results for FAI are less clear, for MAI a strong distinction between the three types of antisolvents can be seen. Type I antisolvents result in a well-mixed and clear solution, while for type II, the solvents are well mixed but a significant amount of white precipitates are present. For type III, a liquid phase separation in combination with a yellowish color change is observed.

Armed with these results, we can understand the mechanisms leading to the trends discussed above. In a simplistic model, the application of the antisolvent triggers two distinct processes. Adding the liquid antisolvent on top of the thinned precursor solution extracts the DMF:DMSO from the underlying layer via diffusion. During this period, the constant supply of neat antisolvent over the precursor layer maintains a high gradient for diffusion and makes this process very efficient. The extraction of the solvent from the precursor solution simultaneously triggers nucleation and solidification of the perovskite material, resulting in polycrystalline films as observed. However, precursor molecules may also diffuse into the antisolvent layer, in addition to the solvent molecules. The relative differences in the interactions between solvent and precursor material with the given antisolvent determines the effectiveness of the antisolvent treatment at removing the DMF:DMSO while preserving the perovskite composition.

For type I antisolvents, we observe increasing solubility of the organic precursors as the alkyl chain length of the alcohol decreases, as well as an overall high miscibility with the DMF:DMSO solvent mixture (supplementary note S7, Figure S7 b,c). Concurrently, films formed by this type possess a large amount of residual $PbI_2$ when compared to films formed by the other antisolvent types. This suggests that, along with the DMF:DMSO, a considerable amount of the organic halides are also removed by these antisolvents, damaging the film microstructure and leaving behind $PbI_2$ which cannot convert to perovskite, as the stoichiometry has been irrevocably altered. This is consistent with the use of methanol (MeOH) as the antisolvent: the solubility of the organic precursors in MeOH is so high that they are completely removed from the film during the antisolvent application, yielding only a yellow film of residual $PbI_2$ (supplementary note S8, Figure S8). Due to the differences in diffusion rates, short application times are still sufficient to remove the DMF:DMSO, but ineffective at removing the organic halides, thus yielding high quality films with a good PV performance.

The type II antisolvents have the ideal combination of properties. They exhibit low solubility for the organics, as indicated by the amount of precipitates seen in Figure 6, but are still miscible with the DMF:DMSO host solvent and therefore provide effective removal of the solvent mixture. Due to the large mismatch in diffusion rates for these antisolvents, the duration of the antisolvent application is largely irrelevent, as they will only act to remove the DMF:DMSO while leaving the perovskite composition intact. However, as shown in Figure 2, certain type II antisolvents still display a difference between fast and slow performance. This is caused by their low, but non-negligible, solubility for the organic iodides – EA, for example, possesses the smallest amount of precipitates of all the type II antisolvents, and also has the largest performance difference, indicating that it sits somewhere between type I and II. This is the reason for EA's inclusion in type II: despite having similar PV behavior to BuOH, the solubility test shows a clear distinction between the two.

As noted previously, the type III antisolvents often possess poor film coverage when formed via a fast antisolvent application. This is likely caused by the immiscibility of the solvents, indicated by the liquid phase separation observed in the top panel of Figure 6. When applied quickly, there is inadequate time for the DMF:DMSO to diffuse across the liquid-liquid interface into the antisolvent, and the film coverage suffers as a result, analogous to film fabrication without any antisolvent. Only in the very center, where the film is most exposed to the antisolvent, is the host solvent effectively removed and a perovskite crystal phase formed. Prolonging contact between the antisolvent and precursor layers allows the relatively slow diffusion process enough time to proceed to completion. Indeed, films fabricated via a slow antisolvent application, which is sufficient to extract the host DMF:DMSO solvents, result in high quality perovskite films with excellent photovoltaic performance. Figure 6 summarizes the mechanisms involved in perovskite film formation by the different categories of antisolvents.

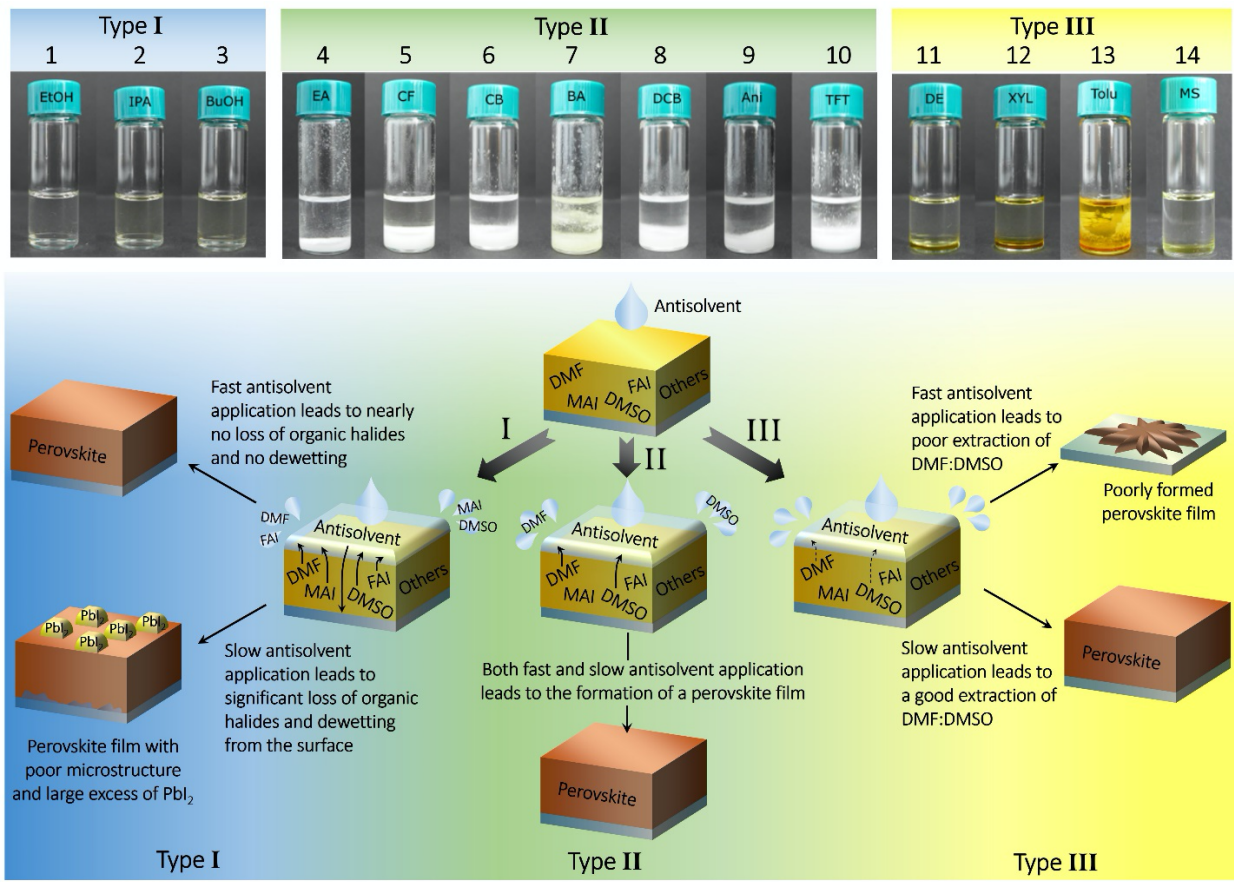

**Figure 6.** Top: solubility of MAI in a solution of DMF:DMSO:antisolvent, meant to simulate the perovskite film intermediate phase during the antisolvent step of fabrication. Bottom: summary of the various mechanisms involved in perovskite film formation by the different categories of antisolvents.

Recently, significant efforts have been devoted to investigating MA-free perovskite compositions with impressive device performance demonstrated by a range of methods.[29–31] We note that while the focus of this work were photovoltaic devices with a triple-cation perovskite composition, we observe that tuning the duration of antisolvent application is similarly critical also in the MA-free devices, (supplementary note S10, Figure S10) owing to the presence of FAI in these devices and its similar behavior to what has been discussed here. We note that while the assignment of each antisolvent into a particular category might differ from that described here for the MA containing perovskite compositions, these observations confirm the broader applicability of our approach also for other perovskite solar cells fabricated by the solvent engineering method.

**Role of precursor solution stoichiometry**

Various groups have reported an increase in performance and reproducibility when an excess of PbI$_2$ was included in the precursor solution.[32,33] This partially stems from the fact that the vast majority of research groups treat Cs as an additive rather than a component in their stoichiometry calculations, i.e. setting the ratio of (FA+MA):Pb equal to one. In our case, we include Cs as a component and thus (Cs+MA+FA):Pb = 1; this difference in methodology results in what others would consider a 5% lead excess. For reference, Saliba et al.[34] reported best results with an 8% excess (as calculated without considering Cesium). While some reports suggest that reducing any residual PbI$_2$ is indicative of a superior performing film,[34,35] other reports have found their highest performing devices contain more PbI$_2$ than the controls, leaving the role of PbI$_2$ unresolved.[29,36,37] It is noteworthy that excess PbI$_2$ has been shown to lead to a reduction in device stability.[38,39] Because the results of our examination of the effect of antisolvent extrusion rate in this study revealed alterations to the stoichiometry, microstructure, and residual PbI$_2$ in our triple cation devices, we also examined how the 'fast' vs. 'slow' antisolvent application interacts with variations in the initial precursor stoichiometry.

To investigate how changing the duration of the antisolvent application step interacts with variations in the precursor solution stoichiometry, we measured the performance of photovoltaic devices fabricated with a varying initial amount of organic precursors, i.e. Cs$_{0.05}$(FA$_{0.83}$MA$_{0.17}$)$_{0.95}$·**x**Pb(I$_{0.9}$Br$_{0.1}$)$_3$ with **x** ranging from 0.9 to 1.1. We selected anisole, which as a type II antisolvent reliably produces high performance devices with very little difference between fast and slow antisolvent application.

Figure 7 shows that for both antisolvent applications the PCE remains largely unaffected by the deliberate introduction of a deficiency of organic precursors, with only the films that are 10% deficient in organic iodides (x=0.9) resulting in a lower performance. On the other hand, excess of organic iodides leads to a significant difference in the performance of devices fabricated via the slow and fast antisolvent application. In the case of a fast application, the devices maintain their performance up to 6% excess, and show only a moderate reduction in performance for larger x values. Contrary to this, slow application causes a loss of performance already at 3% excess of organic iodides and results in a severe loss of efficiency for higher excess values. Note that, for easier visualization due to a significant amount of overlapping data, Figure 7 only displays the top 10 devices for each category. However, the trend is identical for all devices, and for the interested reader, a full display of all devices' performance is shown in supplementary note S11, Figure S11. These results suggest that the common use of excess lead iodide (i.e. deficiency in organic cations) does not necessarily originate solely from an innate advantage in terms of device efficiency, as equally efficient devices can be fabricated with a similar excess of organic cations, provided the right duration of the antisolvent application is adopted. Instead, it is likely that the use of lead iodide excess is related to the enhanced reproducibility of the performance of solar cells fabricated by different researchers (with naturally differing rates of antisolvent application) in this stoichiometry regime. Importantly, we observe that the highest performance is achieved for fully stoichiometric devices (x=1, PCE=22.4%), thus eliminating the need to introduce excess $PbI_2$ into the precursor solution.

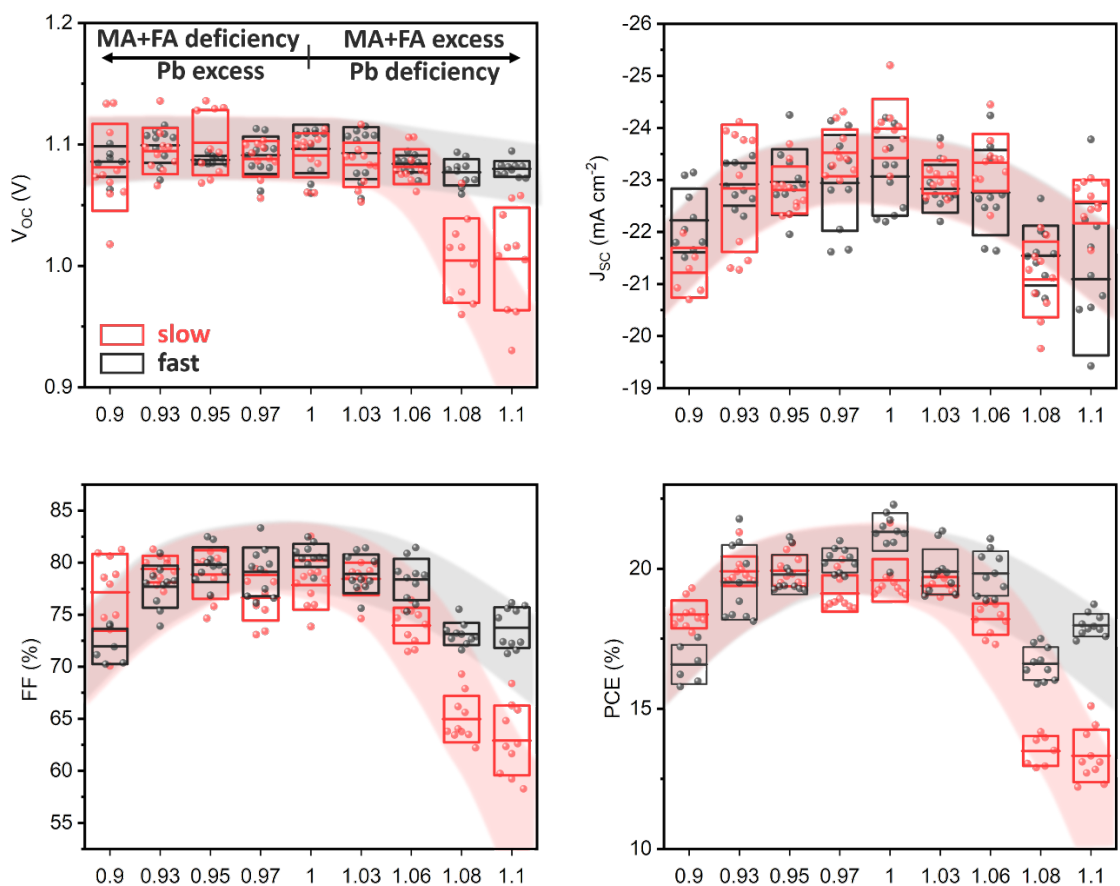

**Figure 7**. PV performance for triple cation perovskite solar cells with varying stoichiometric ratios, for both 'fast' and 'slow' antisolvent speeds. Here, [MA+FA]$_1$ is the amount of organics in an exactly stoichiometric precursor solution. For clarity, only the top 10 devices for each category are shown. The red and black shaded regions are shown as a guide for the eye.

**Conclusion**

In summary, we present a simple approach to the fabrication of high efficiency perovskite photovoltaic devices from any antisolvent. We demonstrate that antisolvents can be categorized in three groups depending on two factors: (1) their ability to dissolve the organic precursor components and (2) their miscibility with the host perovskite solution solvents. These two factors dictate the optimal application time for each antisolvent, allowing to form high quality perovskite layers and efficient photovoltaic devices from any antisolvent. Moreover, we demonstrate that by employing this optimal antisolvent application time, high-efficiency devices can be made from a broad range of precursor stoichiometries, tolerating both excess and deficiency of organic iodides by up to 6%. These results not only enhance the fundamental understanding of the role of antisolvents in the film formation of perovskite solar cells, but also provide a simple route to achieve high-efficiency devices with increased reproducibility.

**Methods**

### Materials

Glass substrates, pre-coated with a central ITO stripe, were purchased from PsiOTec Ltd. Methylammonium iodide (MAI, $CH_3NH_3I$) and formamidinium iodide (FAI, $HC(NH_2)_2$) were purchased from GreatCell Solar. $PbI_2$ and $PbBr_2$ were purchased from TCI. $PC_{60}BM$ was purchased from Solenne BV. All other materials and solvents were purchased from Sigma-Aldrich. All materials were stored in a nitrogen-filled glovebox and used as received.

### Solution preparation

Perovskite films were fabricated from precursor solutions using the recipe $Cs_{0.05}(FA_{0.87}MA_{0.13})_{0.95}Pb(I_{0.9}Br_{0.1})_3$. We used a sequential solution method to prepare exactly stoichiometric 1 M precursor solutions. To do so, 2 M solutions of CsI, $PbI_2$, and $PbBr_2$ were prepared by dissolving each in a 4:1 (v/v) mixture of DMF:DMSO and heating at 180°C, and CsI in pure DMSO at 150°C. After being dissolved, the volume of these solutions expands due to the presence of the solute, thus their "true" concentration will be lower than what is found by simply dividing the molecular weight ($M_w$) of the powder by the volume of the solvent added. To determine this true concentration, the mass of a known volume of each solution was measured, from which the molarity can be calculated. Once the true concentration was known, each solution was diluted by adding the appropriate solvent until the desired concentration of 1.155M was reached, and then the CsI, PbI$_2$, and PbBr$_2$ solutions were mixed in a volume ratio of 0.05:0.85:0.15, yielding a 1.1M solution of Cs$_{0.05}$Pb(I$_{1.75}$Br$_{0.3}$), which we term the inorganic stock solution. In two separate vials FAI and MAI powders were added and weighed, into which the appropriate amount (0.95:1 molar ratio) of inorganic stock was added. This creates two new solutions, of the formula Cs$_{0.05}$(FA or MA)$_{0.95}$Pb(I$_{0.9}$Br$_{0.1}$)$_3$. Finally, these two solutions were mixed in a 5:1 v/v ratio, in order to achieve the final molecular formula Cs$_{0.05}$(FA$_{0.83}$MA$_{0.17}$)$_{0.95}$Pb(I$_{0.9}$Br$_{0.1}$)$_3$. For the MA-free devices, the last step was omitted to yield a precursor solution of Cs$_{0.05}$(FA)$_{0.95}$Pb(I$_{0.9}$Br$_{0.1}$)$_3$.

### Device fabrication

PV devices were fabricated in the device stack Glass/ITO/PTAA/PFN/CsFAMA/PCBM/BCP/Ag. First, Glass/ITO substrates were sequentially cleaned by sonication in 2% Hellmanex detergent, deionized water, acetone, and isopropyl alcohol. After being blown dry, the substrates were exposed to an oxygen plasma at 100 mW for 10 min to remove any residual contamination. Immediately after plasma cleaning, the devices were transferred to a drybox (<2% relative humidity), where poly(triaryl amine) (PTAA) was spin coated from a 1.5 mg/mL solution in toluene at 2000 RPM for 30 s, followed by a 10 min annealing step at 100°C. After letting the substrates cool for 5 min, PFN was then spin coated from a 0.5 mg/mL solution in anhydrous methanol at 5000 RPM for 30 s, with no thermal annealing. The perovskite layer was spin coated with a 2-step recipe, first at 1000 RPM for 10 s followed by 5000 RPM for 30s. 200 uL of antisolvent was dripped onto the substrate 5 s as a continuous droplet before the end of the second step, at either a fast or a slow rate. The as-spun samples were annealed at 100 °C for 30 min and then transferred to a nitrogen-filled glovebox.

For the electron transport side, phenyl-C61-butyric acid methyl ester (PCBM) was spin coated dynamically from a 20 mg/mL solution in chlorobenzene at 2000 RPM for 30s, followed by a 10 min. anneal at 100°C. After letting the substrates cool for 5 min, bathocuproine (BCP) was spin coated dynamically from a 0.5 mg/mL solution in IPA at 4000 RPM for 30s, with a 5 min. anneal at 70°C. To complete the devices, the samples were then transferred, without breaking the inert atmosphere, to a thermal evaporator where 80 nm silver electrodes were deposited at an initial rate of 0.01 nm/s for the first 15 nm, then 0.1 nm/s for the remainder.

### Adjusting the precursor stoichiometry

During the precursor solution preparation, an inorganic stock solution was created consisting of $Cs_{0.05}Pb(I_{1.75}Br_{0.3})$, which was then added to MAI and FAI powders to create the final perovskite solution $Cs_{0.05}(FA_{0.83}MA_{0.17})_{0.95}Pb(I_{2.7}Br_{0.3})$. To adjust the [MAI+FAI]:Pb ratio, a smaller volume of inorganic stock solution was added instead of the exact amount, creating an (organic) overstoichiometric perovskite solution of a known volume. This solution was then used as above to fabricate a number of devices, and the resulting solution volume calculated. Then, to incrementally adjust the stoichiometry, a small volume of the inorganic stock was added to the precursor solution to adjust the stoichiometry to the desired ratio. This process was repeated until the full stoichiometry range was covered.

### J-V characterization

Current density-voltage measurements were performed under simulated AM 1.5 light with an intensity of 100 mW/cm$^2$ (Abet Sun 3000 Class AAA Solar Simulator). The intensity was calibrated using a Si reference cell (NIST traceable, VLSI), and corrected by measuring the spectral mismatch between the solar spectrum, reference cell, and the spectral response of the PV device. The mismatch factor obtained was approximately 1.1. Cells were scanned using a Keithley 2450 source measure unit from 1.2 to 0 V and back, with a step size of 0.025 V and a dwell time of 0.1s, after light soaking for 2 s at 1.2V. The pixel area was 3 mm x 1.5 mm.

### X-ray and Ultraviolet Photoemission spectroscopy (XPS/UPS)

Samples for XPS (Glass/ITO/PTAA/PFN/CsFAMA) were prepared as described above and transferred into the ultrahigh vacuum chamber of the XPS system (ThermoScientific ESCALAB 250Xi, Specs PHOIBOS 100) for measurement. All measurements were performed in the dark, and 5 spots per sample were measured and averaged to acquire the statistics. XPS measurements were performed using an XR6 monochromated AlKα source (hv = 1486.6 eV) and a pass energy of 20 eV.

### 2D X-ray Diffraction (2D XRD)

2D XRD measurements of the CsFAMA films on ITO/glass/PTAA/PFN were conducted using a Rigaku SmartLab diffractometer with a 9kW rotating copper anode in Bragg-Brentano geometry. Diffraction patterns (intensity vs. 2θ) were recorded with a HyPix3000 detector, collected at 2°/min in a coupled Θ/2Θ scan from 0-55° and a detector distance of 110mm.

### Scanning Electron Microscopy (SEM)

SEM was performed using a JSM-7610F FEG-SEM (Jeol). Samples were mounted on standard SEM holders using conductive Ag paste to avoid sample charging. Images were recorded using a secondary electron detector (LEI) at an acceleration voltage of 1.5 kV and a chamber pressure <$10^{-6}$ mbar.

**Author contributions**

A. D. T. and Q. S. contributed equally to this work. A. D. T. and Y. V. conceptualized the study. Y.V. supervised the work. A. D. T., Q. S., Q. A., and T. S. fabricated and characterized the photovoltaic devices. Q. S. performed the solubility and miscibility tests. Q. S. and Y. H. performed the XPS measurements and analysis. F. P. performed the SEM and 2D XRD measurements and analysis. K. P. G. quantified the duration of antisolvent application and M. L. performed the optical microscopy imaging. A. D. T., K. P. G., and Y. V. wrote the manuscript which has been commented on and edited by all co-authors.

**Acknowledgments**

The authors would like to kindly thank Prof. U. Bunz for providing access to device fabrication facilities and Prof. J. Zaumseil for access to the SEM and XRD. This project has received funding from the European Research Council (ERC) under the European Union's Horizon 2020 research and innovation programme (ERC Grant Agreement n° 714067, ENERGYMAPS) and the Deutsche Forschungsgemeinschaft (DFG) in the framework of the Special Priority Program (SPP 2196) project PERFECT PVs (#424216076).

# Supporting Information

**A General Approach to High Efficiency Perovskite Solar Cells by Any Antisolvent**

*Alexander D. Taylor,[1,2,3] Qing Sun,[1] Katelyn P. Goetz,[1,2,3] Qingzhi An,[1,2,3] Tim Schramm,[2] Yvonne Hofstetter,[1,2,3] Maximillian Litterst,[1] Fabian Paulus,[1,3] and Yana Vaynzof[1,2,3]\**

[1] Kirchhoff Institute for Physics and Centre for Advanced Materials, Ruprecht-Karls-Universität Heidelberg, Im Neuenheimer Feld 227, 69120 Heidelberg, Germany

[2] Integrated Centre for Applied Physics and Photonic Materials, Technical University of Dresden, Nöthnitzer Str. 61, 01187 Dresden, Germany

[3] Center for Advancing Electronics Dresden (cfaed), Helmholtzstraße 18, 01089 Dresden, Germany

**Corresponding Author**

*Yana Vaynzof, e-mail: yana.vaynzof@tu-dresden.de

# Contents

## S1. Calculation of the Fast and Slow Antisolvent Extrusion Rate

To calculate the antisolvent extrusion rate, five simulated antisolvent applications for both fast and slow were recorded on video using a dark liquid for contrast. Each video was then broken up into individual frames, wherein the frame at which the application started and finished was noted. Using the known frame rate, the duration for each was calculated and the average found. Dividing the volume (200 µL) by this average duration yielded the average rate for 'fast' and 'slow' application.

## S2. Antisolvent Characteristics

| Type | Solvent | Density [g/ml] | Boiling point [°C] | Dipole moment [D] |
|---|---|---|---|---|
| I | EtOH | 0.79 | 78 | 1.69 |
| | IPA | 0.79 | 83 | 1.66 |
| | BuOH | 0.81 | 118 | 1.66 |
| II | EA | 0.90 | 77 | 1.78 |
| | CF | 1.49 | 61 | 1.15 |
| | CB | 1.11 | 131 | 1.69 |
| | BA | 0.88 | 126 | 1.87 |
| | DCB | 1.30 | 180 | 2.50 |
| | Ani | 1.00 | 154 | 2.30 |
| | TFT | 1.19 | 103 | 2.86 |
| III | DEE | 0.71 | 35 | 1.15 |
| | Xyl | 0.86 | 139 | 0.33-0.37 |
| | Tol | 0.87 | 111 | 0.36 |
| | Mesit | 0.86 | 164.7 | 0.047 |

**Table S1**: Physiochemical properties of all 14 solvents. The solvents are abbreviated as follows: EtOH is ethanol, IPA is isopropanol, BuOH is butyl alcohol, EA is ethyl acetate, CF is chloroform, CB is chlorobenzene, BA is butyl acetate, DCB is 1,2-dichlorobenzene, Ani is anisole, TFT is trifluorotoluene, DEE is diethyl ether, Xyl is m-xylene, and Tol is toluene, Mesit is mesitylene.

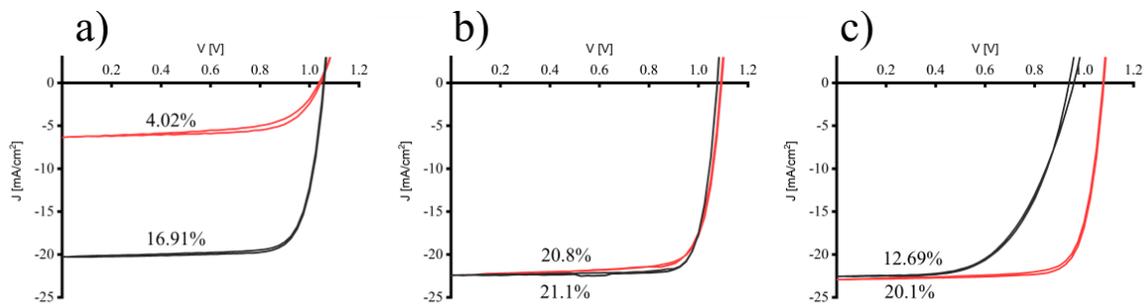

**Figure S1**: Sample JV curves for a type I, II, and III device highlighting the differences between fast and slow antisolvent application. a) ethanol, b) anisole, c) toluene. Red indicates 'slow' and black indicates 'fast'.

## S3. XPS Measurements

Samples for X-ray photoemission spectroscopy (XPS) measurements were prepared as described above on PFN/PTAA coated ITO substrates and then tranferred to the ultrahigh vacuum chamber of the XPS system (Thermo Scientific ESCALAB 250Xi). XPS measurements were performed using a XR6 monochromated Al Kα source (hν = 1486.6 eV) and a pass energy of 20 eV.

The iodine to lead atomic ratio (I/Pb) is calculated by dividing the atomic percentage of iodine by that of lead as obtained from the collected Pb4f and I3d spectra. The bromide to lead (Br/Pb), caesium to lead (Cs/Pb), FA to lead (FA/Pb) and MA to lead (MA/Pb) atomic ratios are obtaied in a similar way. The N1s spectrum of the triple cation perovskite exhibts two peaks at around 400.6 eV and 402.9 eV, representing the FA and MA, respectively (Figure S2a). The atomic percentages of the FA and MA are quantified by fitting the corresponding N1s peaks.

Note that the large variation in the MA/Pb among different acquisition spots for each single sample (Figure S2b) is the result of the fitting, because the N1s peak assigned to MA is low and noisy.

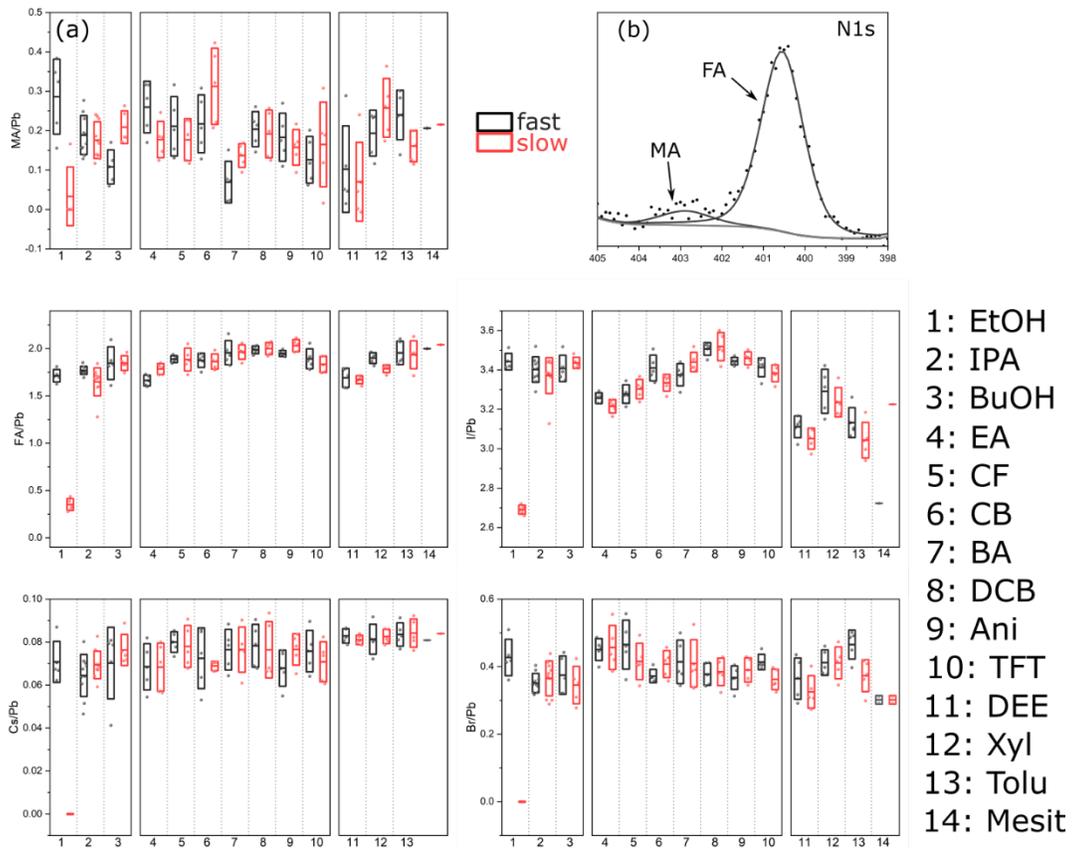

**Figure S2**: XPS characterization of 'fast' vs. 'slow' devices. a) example N1s peaks, showing the signal separation of nitrogen peaks from MA and FA. b) Atomic ratios between lead and the other constituents (MA, FA, I, Br, Cs) of triple cation perovskite.

## S4. SEM and 2D-XRD for Type II Antisolvents

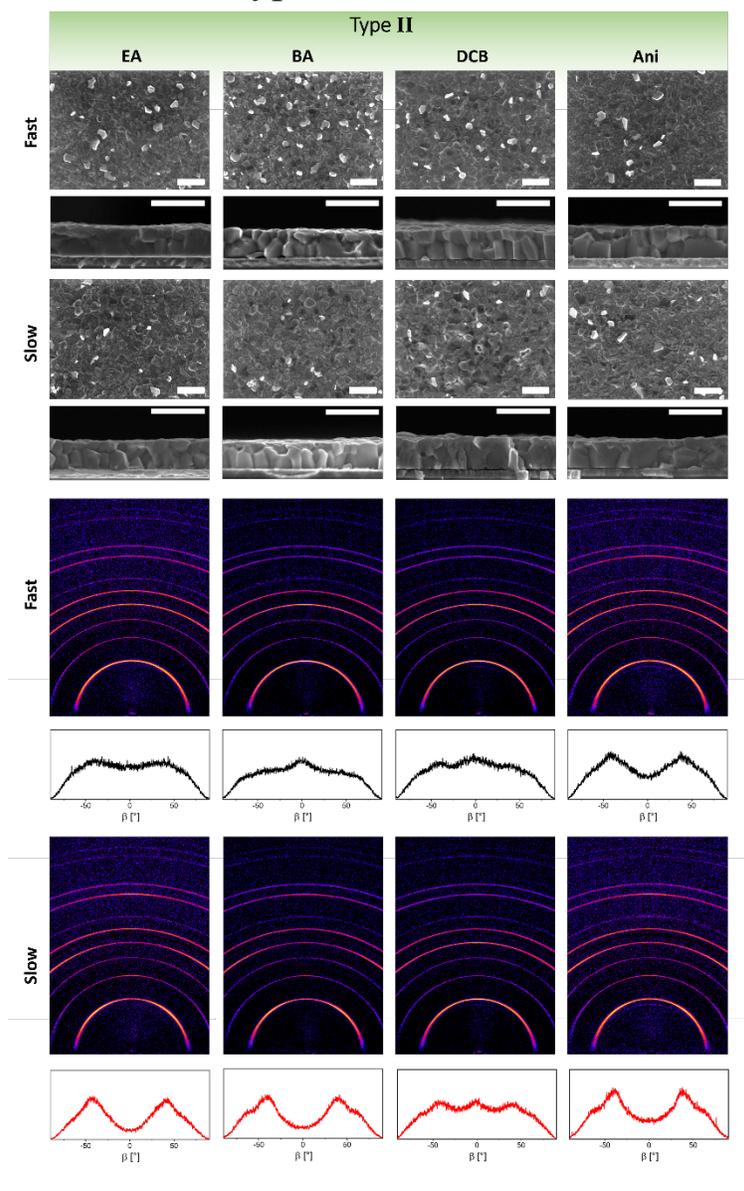

**Figure S3**: Top: surface and cross-sectional scanning electron microscopy images of perovskite films formed from selected type II antisolvents (EA, BA, DCB and Ani). Scale bar is 1 µm. Bottom: 2D XRD maps and corresponding ß integration of the (100) reflection to visualize changes in grain orientation.

## S5. Optical Images of Type II and III Antisolvent Treatments

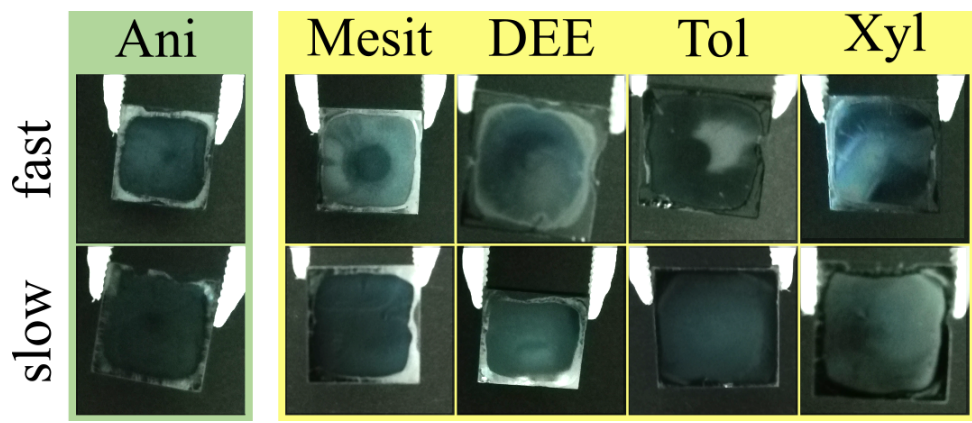

**Figure S4**: Photographs of films fabricated using type III antisolvents, with a type II antisolvent (anisole) for reference, fabricated 'fast' and 'slow'. Note that the black/white contrast has been adjusted in order to more easily visualize differences between the films.

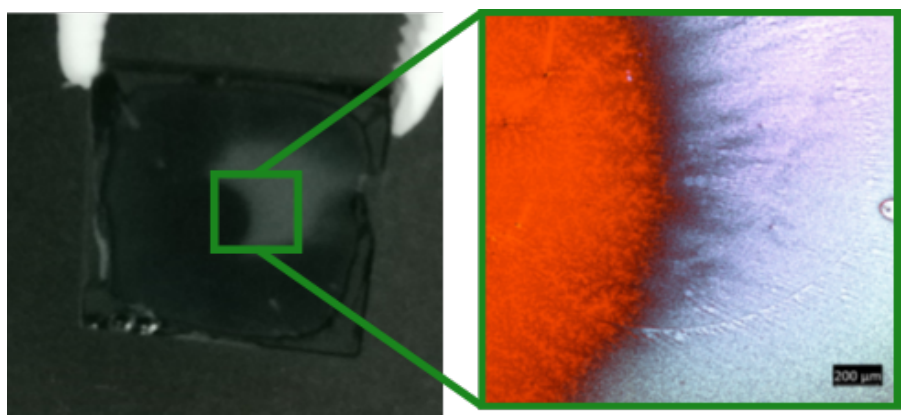

**Figure S5**: Photograph of a perovskite film formed using toluene/fast, and the corresponding optical transmission microscopy image of the highlighted region. While the center circle is the high-quality perovskite seen via SEM, outside the circle is amorphous material.

## S6. SEM and 2D-XRD for Type III Antisolvent Mesitylene

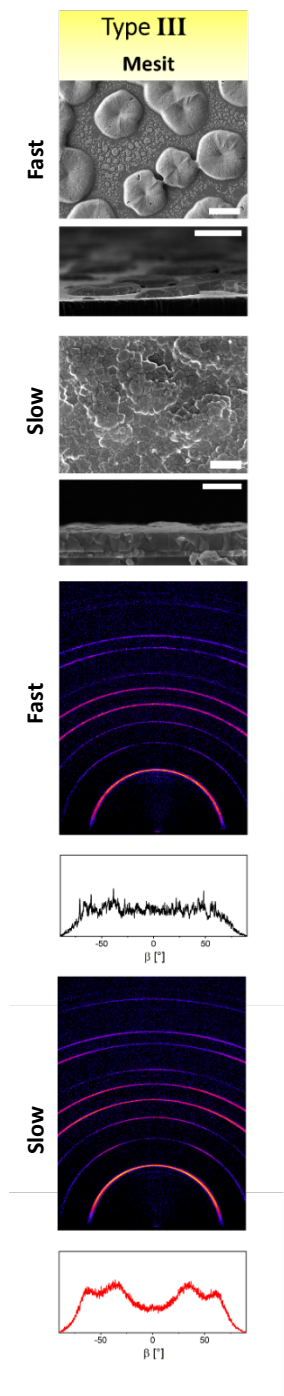

**Figure S6**: Top: surface and cross-sectional scanning electron microscopy images of perovskite films formed from type III antisolvent Mesit. Scale bar for fast antisolvent application is 10 μm and for slow antisolvent application is 1 μm. Bottom: 2D XRD maps and corresponding ß integration of the (100) reflection to visualize changes in grain orientation.

## S7. Solubility of FAI in DMF:DMSO and Antisolvents

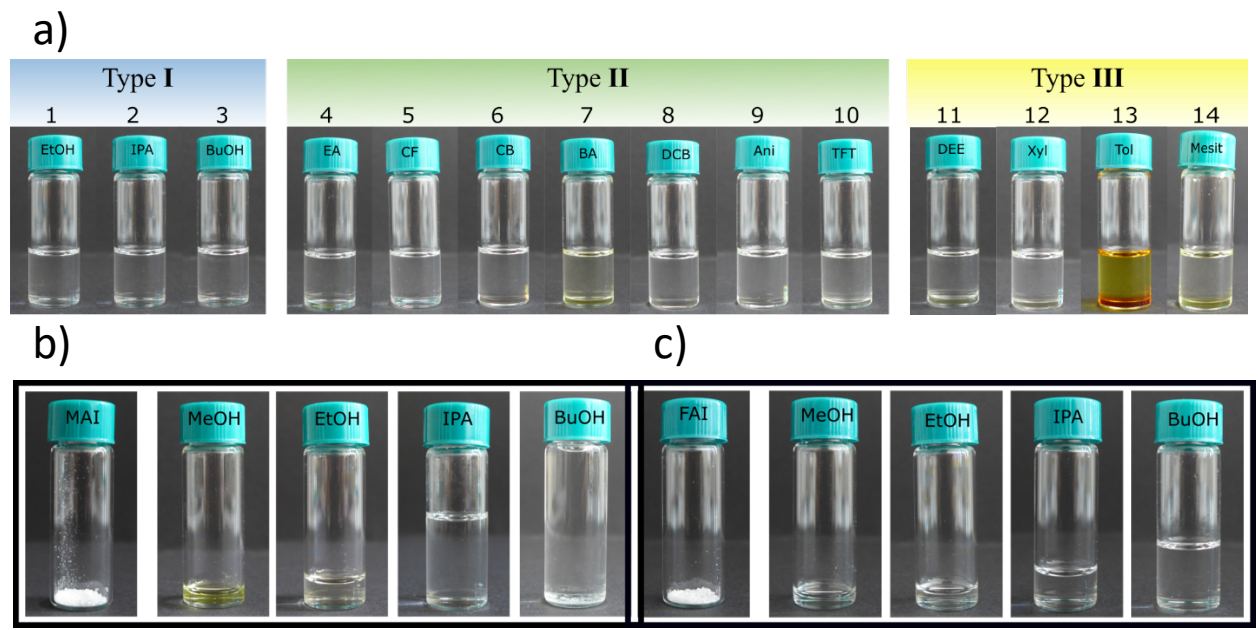

**Figure S7**: a) Solubility of FAI in a solution of DMF:DMSO:antisolvent, meant to simulate the perovskite film intermediate phase during the antisolvent step of fabrication. b,c) Amount of solvent (alcohols) required to fully dissolve 100 mg of MAI* and FAI, respectively. *Note that only the MAI in the BuOH solution is not fully dissolved.

## S8. Optical Image of Film with Methanol as Antisolvent

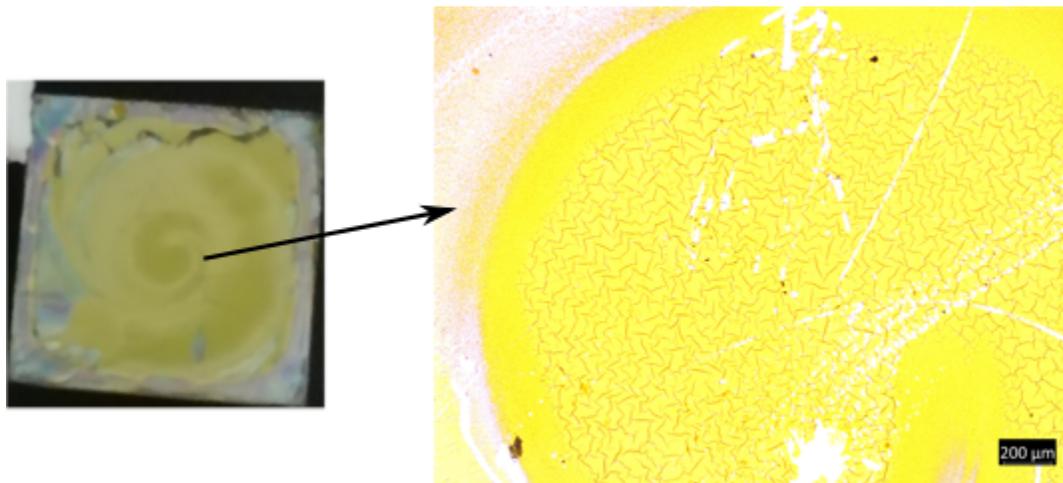

**Figure S8**: Film resulting from the use of methanol as antisolvent, leaving only residual $PbI_2$.

## S9. UV-Vis of Type III antisolvents

A second phenomenon displayed by the type III solvents is the yellowish color change. UV-vis absorption measurements of the same 2 M MAI/FAI solutions in DMF:DMSO mixed with the type II antisolvents reveal absorption onsets and peaks consistent with iodide ($I_2$), as well as more reduced iodine species such as $I_3^-$, shown in Figure S9. These results indicate the possible degradation of the MAI/FAI molecules in solution. However, we note that this process is much slower than the duration during antisolvent application and is thus unlikely to affect the resulting film.

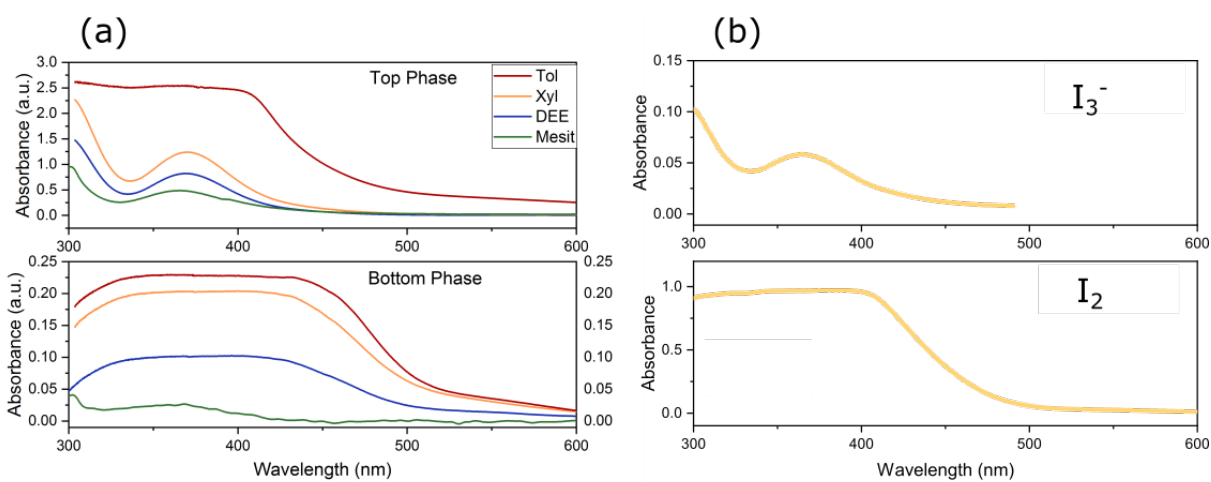

**Figure S9**: a) UV-vis extinction spectra of 2 M MAI/FAI solutions, dissolved in 6:1 antisolvent:[DMF:DMSO], using the Type III antisolvents for the top and bottom separated liquid phases. b) Sample extinction spectra of $I_2$ and $I_3^-$ for reference.[1]

## S10. MA-Free Perovskite with Type I, II, and III Antisolvents

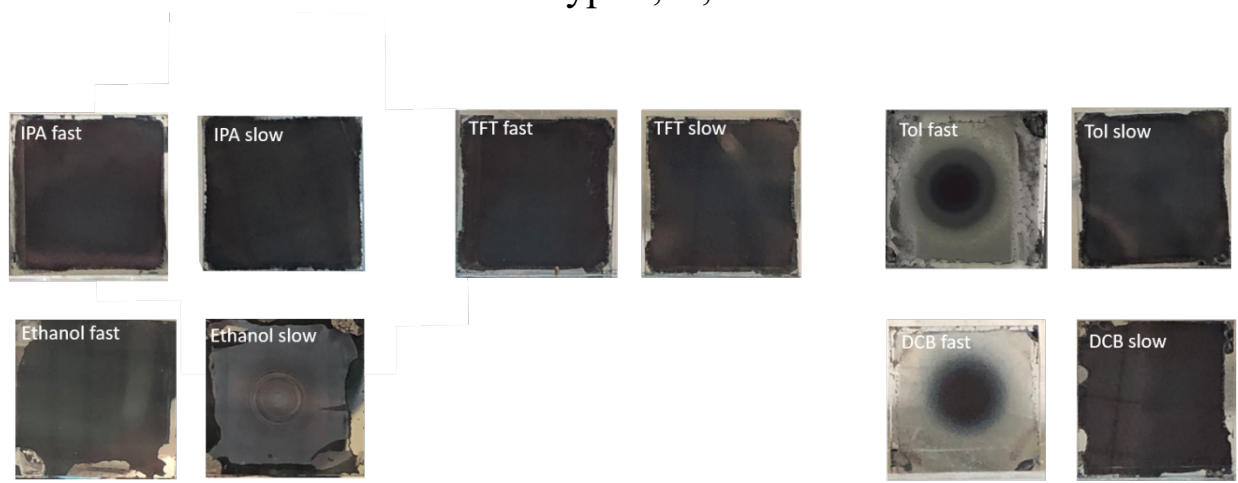

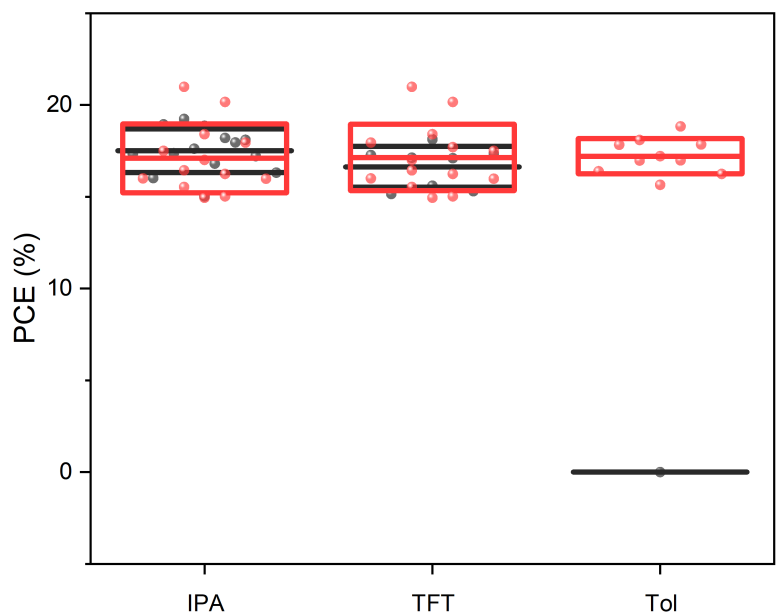

**Figure S10**: Photographs and PV data of perovskite PV devices fabricated using an MA-free composition of $Cs_{0.1}FA_{0.9}PbI_{2.9}Br_{0.1}$, with isopropyl alcohol (type I), trifluorotoluene (type II), and toluene (type III) antisolvents applied fast and slow.

## S11. Full PV Characteristics as a Function of Stoichiometry

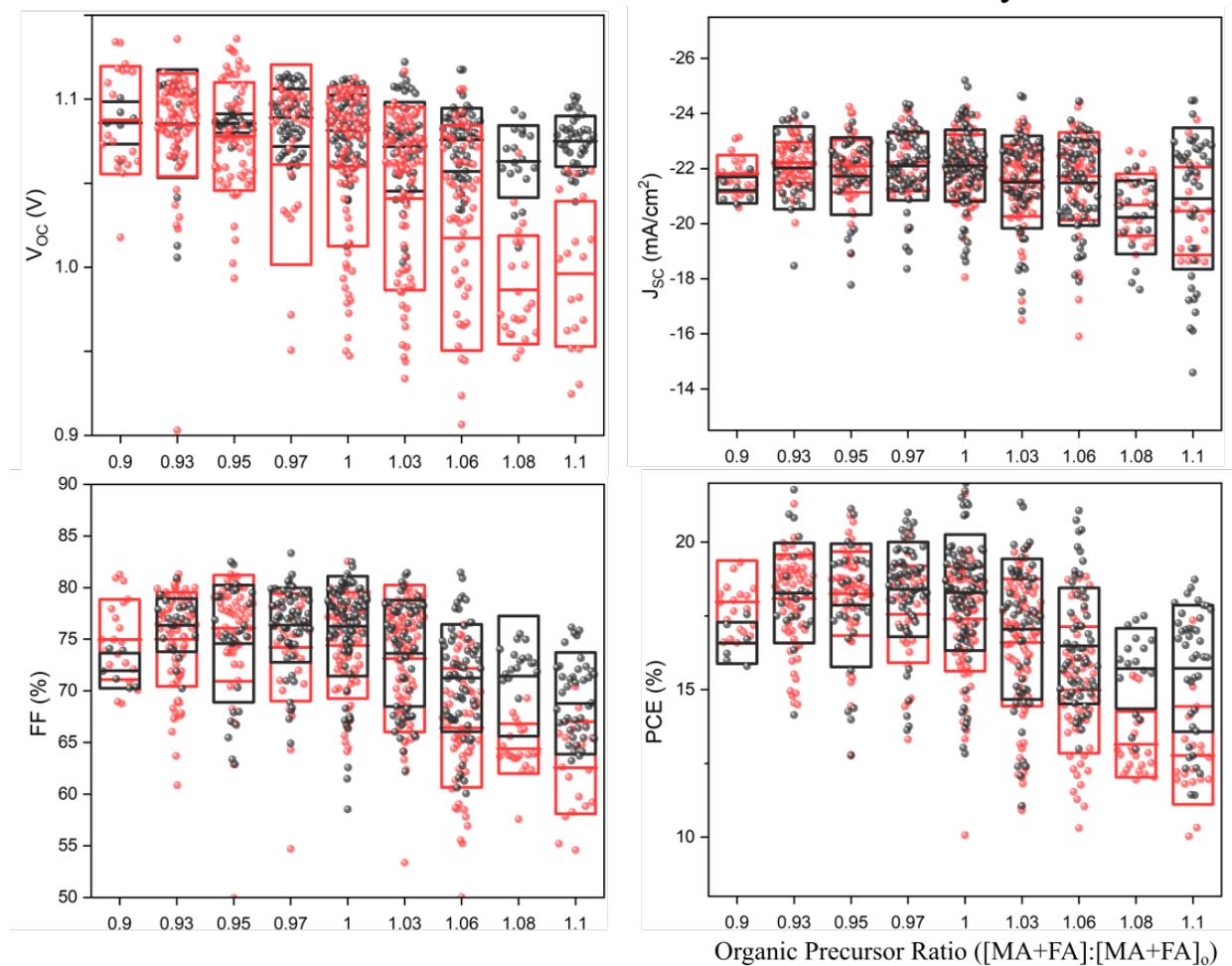

**Figure S11**: Figure 7 from the main text, with all of the data displayed instead of only the top 10 pixels. Note that the overall trends remain identical, but are simply more difficult to follow due to the large number of overlapping points.